\documentclass[12pt]{article}%
\usepackage{citesort}
\usepackage{amsmath}
\usepackage{graphicx}%
\usepackage{amsfonts}%
\usepackage{amssymb}
\begin{document}

\begin{center}

\textbf{{\large Exclusive Higgs and dijet production by double pomeron
exchange.}}

\textbf{{\large The CDF upper limits}}
\end{center}

\vskip                                3mm

\begin{center}
Adam Bzdak
\end{center}

\vskip                                3mm

\begin{center}
M. Smoluchowski Institute of Physics, Jagiellonian University \\
Reymonta 4, 30-059 Krak\'{o}w, Poland

E-mail: \texttt{bzdak@th.if.uj.edu.pl}
\end{center}

\vskip      .5cm

\begin{quotation}
We use as a starting point the original, central inclusive Bialas-Landshoff
model for Higgs and dijet production by a double pomeron exchange in
\textit{pp }(\textit{p\={p}}) collisions. Next we propose the simple extension
of this model to the exclusive processes. We find the extended model to be
consistent with the CDF Run I, II upper limits for double diffractive
exclusive dijet production. The predictions for the exclusive Higgs boson
production cross sections at the Tevatron and the LHC energies are also presented.

\bigskip

\end{quotation}

\section{Introduction}

The discovery of the Higgs boson is one of the main goals of searches at the
present and next hadronic colliders, the Tevatron and the LHC.

One appealing production mode, the double pomeron exchange (DPE) one, was
proposed some time ago in Refs. \cite{Schafer,Bial-Land}. In the following
papers this subject was discussed from different perspectives
\cite{Cudell,Levin,Khoze-Higgs-Jets,Khoze-Higgs,ET4,Cox,Peschan-G,Peschan,Ingelman}%
. Despite some progress the serious uncertainties are still present that do
not allow to get fully reliable predictions needed for future experiments.
This reflects our present limitted understanding of the nature of the
diffractive (pomeron) reactions.

The best way to reduce these uncertainties is to study other double pomeron
exchange processes and compare them with existing data. A particularly
enlightening process is the DPE production of two jets (dijets). Such a
process was originally discussed at the Born level in \cite{Barera}. Later the
dijet production was studied in \cite{Khoze-Higgs-Jets,Koze-Jets} and in
\cite{Pumplin,Bial-Szer-Janik,Cox,Peschan-G,Peschan,Ingelman,Forshaw-jets,Brazylia,Bzdak,Bzdak2}%
.

One generally considers two types of DPE events when colliding hadrons remain
intact, namely exclusive and central inclusive one (or central inelastic). In
the exclusive DPE event the central object $H$ is produced alone, separated
from the outgoing hadrons by rapidity gaps:%
\begin{equation}
p\bar{p}\rightarrow p+\text{gap}+H+\text{gap}+\bar{p}.
\end{equation}
In the central inclusive DPE event there is an additional radiation $X$
accompanying the central object $H$:%
\begin{equation}
p\bar{p}\rightarrow p+\text{gap}+HX+\text{gap}+\bar{p}.
\end{equation}

Recently, using the Bialas-Landshoff \cite{Bial-Land} model for central
inclusive double diffractive production the cross-section for gluon jet
production was calculated \cite{Bzdak,Bzdak2}. In this model in some
approximation pomeron exchange corresponds to the exchange of a pair of
non-perturbative gluons which takes place between a pair of colliding quarks
\cite{pomeron}. The obtained results together with those for quark-antiquark
jets calculated some time ago \cite{Bial-Szer-Janik} give the full
cross-section for dijet production in double pomeron exchange reactions. The
model was found \cite{Bzdak2} to give correct order of magnitude for the
measured \cite{CDF-1} central inclusive dijet cross sections.

In this Letter we propose the simple extension of this model to the exclusive
processes. We find the extended model to be consistent with the CDF Run I, II
upper limits \cite{CDF-1,CDF-2} for double diffractive exclusive dijet
production. We also present the predictions for the exclusive Higgs boson
production cross sections at the Tevatron and the LHC
energies.\begin{figure}[h]
\begin{center}
\includegraphics[width=5.5cm,height=5cm]{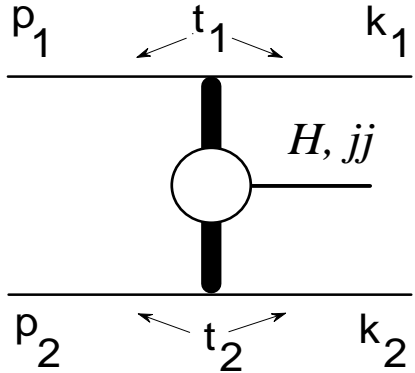}
\end{center}
\caption{Production of Higgs boson $H$, dijet $jj$, by double pomeron
exchange. The colliding hadrons remain intact.}%
\label{ogolny2}%
\end{figure}

\section{Central inclusive dijet production}

\begin{figure}[h]
\begin{center}
\includegraphics[width=3.80cm,height=2cm]{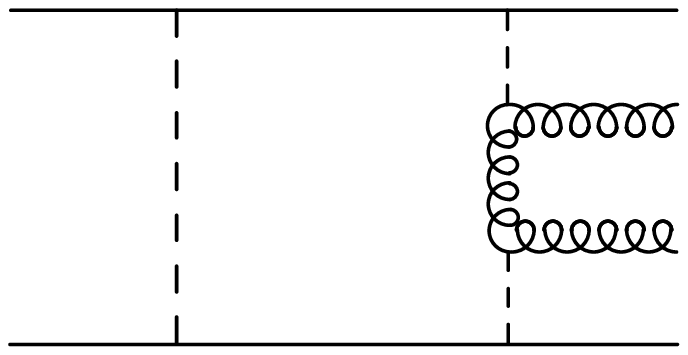}  \hspace{0.3cm}
\includegraphics[width=3.80cm,height=2cm]{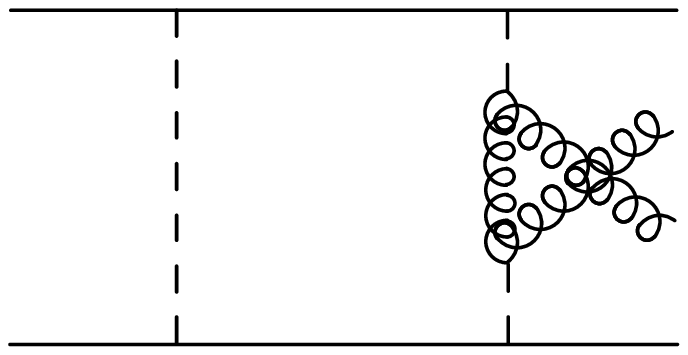}  \hspace{0.3cm}
\includegraphics[width=3.80cm,height=2cm]{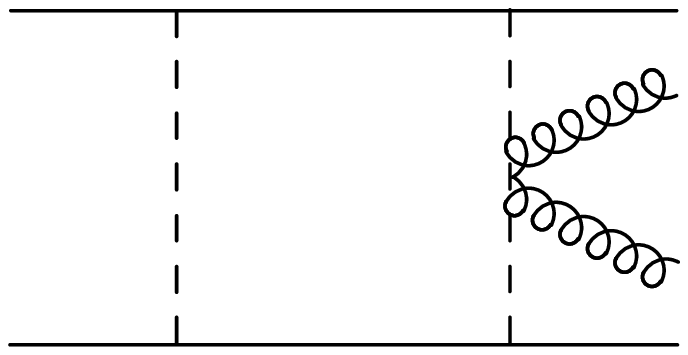}
\end{center}
\caption{Three diagrams contributing to the amplitude of the process of gluon
pair production by double pomeron exchange. The dashed lines represent the
exchange of the non-perturbative gluons.}%
\label{ok}%
\end{figure}

The matrix element for two gluon jet production in the Bialas-Landshoff model
is given \cite{Bzdak} by the $s$-channel discontinuity of the diagrams shown
in Fig. \ref{ok}.

\noindent The square of the matrix element (averaged and summed over spins and
polarizations) is of the form:%
\begin{equation}
\overline{\left|  \mathcal{M}_{pp}\right|  ^{2}}=81\overline{\left|
\mathcal{M}_{qq}\right|  ^{2}}\left[  F\left(  t_{1}\right)  F\left(
t_{2}\right)  \right]  ^{2}, \label{pp}%
\end{equation}
where $\overline{\left|  \mathcal{M}_{qq}\right|  ^{2}}$ is the production
amplitude squared for colliding quarks\footnote{This formula is only valid in
the limit of $\delta_{1,2}<<1$ and for small momentum transfer between initial
and final quarks.}:%
\begin{equation}
\overline{\left|  \mathcal{M}_{qq}\right|  ^{2}}=C\frac{s^{2}}{\left(
u_{_{1}}\right)  ^{2}\left(  u_{_{2}}\right)  ^{2}}\delta_{1}^{2-2\alpha
(t_{1})}\delta_{2}^{2-2\alpha(t_{2})}\exp\left(  2\beta\left(  t_{1}%
+t_{2}\right)  \right)  R^{2}. \label{qq}%
\end{equation}
Transverse momenta of the produced gluons are denoted by $u_{_{1}}$ and
$u_{_{2}}$. The constants $C$ and $R$ will be defined later. $\alpha\left(
t\right)  =1+\epsilon+\alpha^{\prime}t$ is the pomeron Regge trajectory with
$\epsilon\approx0.08,$ $\alpha^{\prime}=0.25$ GeV$^{-2}$ ($t_{1},$ $t_{2}$ are
defined in Fig. \ref{ogolny2}). $F\left(  t\right)  $ = $\exp(\lambda t)$ is
the nucleon form-factor with $\lambda=$ $2$ GeV$^{-2}$. $\delta_{1},$
$\delta_{2}$ are defined as $\delta_{1,2}\equiv1-k_{1,2}/p_{1,2}$ ($k_{1},$
$k_{2},$ $p_{1},$ $p_{2},$ are defined in Fig. \ref{ogolny2}). The factor
$\exp\left(  2\beta\left(  t_{1}+t_{2}\right)  \right)  $ with $\beta$ $=$ $1$
GeV$^{-2}$ \cite{Factor} takes into account the effect of the momentum
transfer dependence of the non-perturbative gluon propagator given by ($p^{2}$
is the Lorentz square of the momentum carried by the non-perturbative gluon):
\begin{equation}
D\left(  p^{2}\right)  =D_{0}\exp\left(  p^{2}/\tau^{2}\right)  .
\label{propag}%
\end{equation}
The constants $C$ and $R$ are defined as:%
\begin{align}
C  &  =\frac{1}{\left(  27\pi\right)  ^{2}}\left(  D_{0}G^{2}\tau\right)
^{6}\tau^{2}\left(  \frac{g^{2}/4\pi}{G^{2}/4\pi}\right)  ^{2},\label{consP}\\
R  &  =9\int d\vec{Q}_{\intercal}^{2}\text{\thinspace}\vec{Q}_{\intercal}%
^{2}\exp\left(  -3\vec{Q}_{\intercal}^{2}\right)  =1. \label{constR}%
\end{align}

Here $G$ and $g$ are the non-perturbative and perturbative quark gluon
couplings respectively\footnote{One should note that the non-perturbative
quark gluon coupling $G$ does not depend on the scale of the process.}. $\tau$
is the range of the non-perturbative gluon propagator (\ref{propag}) and
$D_{0}$ its magnitude at vanishing momentum transfer. From data on the elastic
scattering of hadrons one infers $D_{0}G^{2}\tau=30$ GeV$^{-1}$ and $\tau=1$ GeV.

The constant $R$ reflects the structure of the loop integral. $Q_{\intercal}$
is the transverse momentum carried by each of the three non-perturbative
gluons. $R$ was shown explicitly in Eq. (\ref{qq}) for the reason which will
become clear in the next section.

Taking into account (\ref{qq}) we obtain the following result for the
differential cross-section \cite{Bzdak2}:
\begin{align}
\frac{d\sigma}{d(E_{\intercal}^{2})d(\Delta y)dy}  &  =R^{2}C_{E}\left(
E_{\intercal}\right)  ^{-4}\left(  \frac{s}{4E_{\intercal}^{2}\cosh
^{2}(\frac{\Delta y}{2})}\right)  ^{2\epsilon}\nonumber\\
&  \times\frac{\pi^{3}/4\alpha^{\prime}{}^{2}}{\left(  (\lambda+\beta
)/\alpha^{\prime}-\ln\left[  2E_{\intercal}\cosh(\dfrac{\Delta y}{2})/\sqrt
{s}\right]  \right)  ^{2}-y^{2}}. \label{end3}%
\end{align}
Here $C_{E}=81C/(16\left(  2\pi\right)  ^{8})$. $E_{\intercal}=$
$|u_{1}|=|u_{2}|$ is the transverse energy of one of the produced gluons.
$\Delta y=y_{1}-y_{2},$ $y=(y_{1}+y_{2})/2$ where $y_{1,2}$ are the rapidities
of the produced gluons. For completeness it is necessary to say that the
rapidities $y_{1,2}$ are connected with $\delta_{1},$ $\delta_{2}$ and
$E_{\intercal}$ in the following way:
\[
\delta_{1}\sqrt{s}=E_{\intercal}\exp\left(  y_{1}\right)  +E_{\intercal}%
\exp\left(  y_{2}\right)  ,
\]%
\begin{equation}
\delta_{2}\sqrt{s}=E_{\intercal}\exp\left(  -y_{1}\right)  +E_{\intercal}%
\exp\left(  -y_{2}\right)  .
\end{equation}

The result (\ref{end3}) does not take gap survival effect ($S_{\text{gap}}%
^{2}$) into account $i.e.$ the probability of the gaps not to be populated by
secondaries produced in the soft rescattering. From
\cite{Khoze-Higgs-Jets,CDF-S2} we expect that for the Tevatron energy it is
about $0.05-0.1$. The factor $S_{\text{gap}}^{2}$ is not a universal number
but it depends on the initial energy and the particular final state.
Theoretical predictions of the survival factor, $S_{\text{gap}}^{2}$, can be
found in Ref. \cite{S2-theory}.

The main uncertainty in the expression (\ref{end3}) is the value of
$G^{2}/4\pi$ (see (\ref{consP}))$.$ It is expected to be \cite{G/4pi} about
$1$ but in fact it should be considered only as an order of magnitude estimate.

Let us now make clear the rather $ad$ $hoc$ nature of many of the assumptions
inherent in the Bialas-Landshoff approach \cite{Bial-Land}. The predictions of
this model depend only weakly on energy ($\sim s^{2\epsilon}$). This is a
consequence of the Regge-like dependence implied by Eq. (\ref{qq}). There are
some controversies if such assumption is justified. In our calculations we
assume the exponential form of the non-perturbative gluon propagator
(\ref{propag}). As was already stated in \cite{Bial-Land} there is no reason
to believe that the true form of $D$ is as simple as this. However, we believe
that this is not a serious objection to our model. In the Bialas-Landshoff
approach the produced object (Higgs, dijet $etc.$) is coupled to the
non-perturbative gluons via the perturbative coupling $g$. It is not clear and
the question of consistency could be addressed. Finally, let us note that
estimates in the present Letter are based on the basis of the pure forward
direction. It was first mentioned in \cite{Pumplin} that such approach may
lead to incorrect results.

Integrating (\ref{end3}) over the CDF Run I kinematical range \cite{CDF-1} for
the central inclusive production of dijets of $E_{\intercal}>7$ GeV we obtain
\cite{Bzdak2} the result to be about $70$ nb (with $G^{2}/4\pi=1$ and no
$S_{\text{gap}}^{2}$), to be compared with the CDF measurement of $43$ nb
($43\pm26$ nb). We thus scale our cross section by a factor of $43/70\approx
0.6$, that is:%
\begin{equation}
\frac{S_{\text{gap}}^{2}(\sqrt{s}=2\text{ TeV})}{\left(  G^{2}/4\pi\right)
^{2}}=0.6. \label{Norma}%
\end{equation}

This completes the summary of \cite{Bzdak} and \cite{Bzdak2}.

\section{Exclusive dijet production -- Sudakov factor}

As was already mentioned the calculation presented in the previous section,
based on the original Bialas-Landshoff model, is a central inclusive one
$i.e.$ the radiation is present in the central region of the rapidity.

In order to describe the exclusive processes one has to forbid this radiation.
To do it we include the Sudakov survival factor $T(Q_{\intercal},\mu)$ inside
the loop integral (\ref{constR}) over $Q_{\intercal}$. The Sudakov factor
$T(Q_{\intercal},\mu)$ is the survival probability that a gluon with
transverse momentum $Q_{\intercal}$ remains untouched in the evolution up to
the hard scale $\mu=M_{gg}/2$ where $M_{gg}$ is the mass of the produced
gluons. The function $T(Q_{\intercal},\mu)$ can be calculated as
\cite{Khoze-Higgs-Jets}:%
\begin{equation}
T(Q_{\intercal},\mu)=\exp\left(  -\int_{\vec{Q}_{\intercal}^{2}}^{\mu^{2}%
}\frac{\alpha_{s}\left(  \vec{k}_{\intercal}^{2}\right)  }{2\pi}\frac{d\vec
{k}_{\intercal}^{2}}{\vec{k}_{\intercal}^{2}}\int_{0}^{1-\Delta}\left[
zP_{gg}\left(  z\right)  +\sum\limits_{q}P_{qg}(z)\right]  dz\right)  .
\label{T-def}%
\end{equation}
Here $\Delta=\left|  k_{\intercal}\right|  /\left(  \mu+\left|  k_{\intercal
}\right|  \right)  $, $P_{gg}\left(  z\right)  $ and $P_{qg}(z)$ (we take
$q=u,d,s,\bar{u},\bar{d},\bar{s}$) are the GLAP spitting functions.
$\alpha_{s}$ is the strong coupling constant\footnote{In the following we take
$\alpha_{s}$ at one loop accuracy $i.e.$ $\alpha_{s}\left(  q^{2}\right)
=(4\pi/\beta_{0})\left(  1/\ln\left(  q^{2}/\Lambda^{2}\right)  \right)  $
with $\beta_{0}=9$ and $\Lambda=200$ MeV.}.

Taking into account the leading-order contributions \cite{Splitting} to the
GLAP splitting functions:%
\begin{align}
P_{gg}\left(  z\right)   &  =6\left[  z/(1-z)+(1-z)/z+z\left(  1-z\right)
+\delta\left(  1-z\right)  \left(  11/2-n_{f}/3\right)  \right]  ,\nonumber\\
P_{qg}\left(  z\right)   &  =\left[  z^{2}+\left(  1-z\right)  ^{2}\right]
/2, \label{Split}%
\end{align}
we obtain:%
\begin{align}
\int_{0}^{1-\Delta}zP_{gg}\left(  z\right)  dz  &  =-11/2+12\Delta-9\Delta
^{2}+4\Delta^{3}-3\Delta^{4}/2-6\ln\Delta,\nonumber\\
\int_{0}^{1-\Delta}P_{qg}\left(  z\right)  dz  &  =1/3-\Delta/2+\Delta
^{2}/2-\Delta^{3}/3.
\end{align}

Now to describe the exclusive processes we use the formula (\ref{end3}) with
$R^{2}=1$ replaced by $\tilde{R}^{2}\left(  \mu\right)  $ where $\tilde
{R}\left(  \mu\right)  $ is defined as\footnote{Notice that $\mu>1.5$ GeV is
required so that $9\int_{\Lambda^{2}}^{\mu^{2}}d\vec{Q}_{\intercal}^{2}%
\,\vec{Q}_{\intercal}^{2}\exp(-3\vec{Q}_{\intercal}^{2})=1$.}:%
\begin{equation}
\tilde{R}\left(  \mu\right)  =9\int_{\Lambda^{2}}^{\mu^{2}}d\vec{Q}%
_{\intercal}^{2}\,\vec{Q}_{\intercal}^{2}\exp\left(  -3\vec{Q}_{\intercal}%
^{2}\right)  T\left(  Q_{\intercal},\mu\right)  . \label{R-tylda}%
\end{equation}
The hard scale $\mu=M_{gg}/2$ can be expressed by $E_{\intercal}$ and $\Delta
y$ in the following way:%
\begin{equation}
\mu=E_{\intercal}\cosh(\frac{\Delta y}{2}). \label{Mgg}%
\end{equation}

Naturally a question of internal consistency arises. Namely, the Sudakov
factor uses perturbative gluons whilst in our calculations the Born amplitude
(\ref{qq}) uses non-perturbative gluons. It is not clear what the
non-perturbative gluon is and the extension of the original Bialas-Landshoff
model to the exclusive processes is not straightforward. We hope that taking
the Sudakov factor in the loop integral into account we obtain an approximate
insight into exclusive processes. It should be emphasized that at present our
calculation is a hybrid of perturbative and non-perturbative ideas.

At the end of this section let us notice that the Sudakov factor (\ref{T-def})
does not depend on the sum of the dijet rapidities $y=\left(  y_{1}%
+y_{2}\right)  /2$. This together with the observation that $y^{2}<<\left(
(\lambda+\beta)/\alpha^{\prime}\right)  ^{2}=144$ and $4E_{\intercal}^{2}%
\cosh^{2}(\frac{\Delta y}{2})/s=\delta_{1}\delta_{2}$ $<<1$ leads to the
conclusion that the differential cross section for DPE exclusive dijets
production very weekly depends on the sum of the dijet rapidity $y$. This
feature agrees with the observation found in Ref. \cite{Koze-Jets}. Moreover
the observed power law $E_{\intercal}^{-6.5}$ $($with $\tilde{R}^{2}\sim
E_{\intercal}^{-2.2})$ is close to the observation of Ref. \cite{ET4} $(\sim
E_{\intercal}^{-7.3})$.

\section{CDF Run I, II upper limits}

The CDF collaboration has presented results on upper limits on exclusive DPE
dijet production cross sections.

At Run I ($\sqrt{s}=1.8$ TeV) \cite{CDF-1} the upper bound for exclusive
dijets production was measured to be $3.7$ nb for the kinematic range of
$0.035<\delta_{2}\equiv\delta_{\bar{p}}<0.095$ and jets of $E_{\intercal}>7$
GeV confined within $-4.2<y<2.4$ and the gap requirement $2.4<y_{\text{gap}%
}<5.9$ on the proton side.

At Run II ($\sqrt{s}=1.96$ TeV) \cite{CDF-2} the upper bound for exclusive
dijets of $E_{\intercal}>10$ GeV [$E_{\intercal}>25$ GeV] was measured to be
$970$ $\pm$ $65$(stat) $\pm$ $272$(syst) pb [$34$ $\pm$ $5$(stat) $\pm$
$10$(syst) pb]. The kinematics is following\footnote{We would like to thank K.
Goulianos for a correspondence about this point.}: $0.03<\delta_{2}%
\equiv\delta_{\bar{p}}<0.1$, jets are confined within $-2.5<y<2.5$, the gap on
the proton side is $3.6<y_{\text{gap}}<7.5$.

It should be noted that in the above experiments the protons were not detected
and the DPE events were enhanced by a rapidity gap requirement on the proton
side\footnote{In principle the result (\ref{end3}) should by multiplied by a
factor $(1-\exp[-2(\lambda+\beta-\alpha^{\prime}\ln\delta_{1})\frac{s\left(
1-\delta_{1}\right)  ^{2}}{\exp\left(  2y_{\text{gap}}^{\max}\right)  }])$
where $y_{\text{gap}}^{\max}$ is the maximum value of the gap. In the present
case, $y_{\text{gap}}^{\max}=5.9$ and $7.5$ for Run I and Run II respectively,
this factor is close to $1$.}.

Integrating\footnote{Note an identical final state particle phase space factor
$\frac{1}{2!}$.} (\ref{end3}) over the appropriate kinematical range we obtain
the results shown in Table \ref{jets}. The running coupling constant
$g^{2}/4\pi,$ appearing in (\ref{consP}), is evaluated at $2E_{\intercal
}^{\min}$ $i.e.$ $0.15$, $0.14$, $0.12$ for $E_{\intercal}^{\min}=7,10,25$ GeV
respectively. The factor $S_{\text{gap}}^{2}/\left(  G^{2}/4\pi\right)  ^{2}$
is taken to be $0.6$.\begin{table}[h]
\begin{center}%
\begin{tabular}
[c]{|c|c|c|}\hline\hline
$%
\begin{array}
[c]{c}%
\text{Transverse}\\
\text{energy}%
\end{array}
$ & $%
\begin{array}
[c]{c}%
\text{CDF}\\
\text{upper limits}%
\end{array}
$ & $%
\begin{array}
[c]{c}%
\text{Model}\\
S_{\text{gap}}^{2}/\left(  G^{2}/4\pi\right)  ^{2}=0.6
\end{array}
$\\\hline
$E_{\intercal}>7$ GeV & $3.7$ [nb] & $1$ [nb]\\\hline
$E_{\intercal}>$ $10$ GeV & $970$ $\pm$ $337$ [pb] & $300$ [pb]\\\hline
$E_{\intercal}>$ $25$ GeV & $34$ $\pm$ $15$ [pb] & $3$ [pb]\\\hline
\end{tabular}
\end{center}
\caption{Comparison of the CDF upper limits for DPE exclusive dijet production
with the results obtained in the presented model.}%
\label{jets}%
\end{table}

As can be seen from Table \ref{jets} the obtained results are quite
satisfactory and encouraging. They are also comparable with those obtained in
\cite{ET7KMR,gama} or \cite{Brazylia}.

\section{Exclusive Higgs production}

\begin{figure}[h]
\begin{center}
\includegraphics[width=3.80cm,height=2cm]{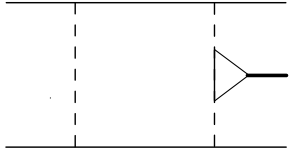}
\end{center}
\caption{The diagram contributing to the amplitude of the process of Higgs
boson production by double pomeron exchange. The Higgs coupling is taken to be
through a t-quark loop. The dashed lines represent the exchange of the
non-perturbative gluons.}%
\label{Higgs_ok}%
\end{figure}

The matrix element for the Higgs production in the Bialas-Landshoff model is
given \cite{Bial-Land} by the $s$-channel discontinuity of the diagram shown
in Fig. \ref{Higgs_ok}. The Higgs coupling is taken to be through a t-quark
loop. 

The square of the matrix element for colliding (anti)protons has the form
\cite{Bial-Land}\footnote{Our result for the matrix element differs from the
result of Bialas and Landshoff by a factor $\exp\left(  2\beta\left(
t_{1}+t_{2}\right)  \right)  $. The missing factor $2$ pointed out in
\cite{Cudell} is also taken into account.}:%
\begin{align}
\overline{\left|  \mathcal{M}_{pp}\right|  ^{2}}  &  =BN^{2}s^{2}\delta
_{1}^{2-2\alpha(t_{1})}\delta_{2}^{2-2\alpha(t_{2})}\left[  F\left(
t_{1}\right)  F\left(  t_{2}\right)  \right]  ^{2}\nonumber\\
&  \times\exp\left(  2\beta\left(  t_{1}+t_{2}\right)  \right)  R^{2}.
\label{H-pp}%
\end{align}
The constant $B$ is defined as:%
\begin{equation}
B=\frac{4\sqrt{2}}{\left(  6\pi\right)  ^{6}}G_{F}\left(  G^{2}D_{0}%
\tau\right)  ^{6}\tau^{2}\left(  \frac{\alpha_{s}}{G^{2}/4\pi}\right)  ^{2},
\label{B}%
\end{equation}
where $G_{F}$ is the Fermi coupling constant and $\alpha_{s}(M_{H})$ is the
perturbative coupling evaluated at a scale $M_{H}$. $N$ is a function of
$M_{t}/M_{H}$. For the Higgs mass $M_{H}<$ $2M_{t}\approx350$ GeV this
function is given by \cite{Schafer}:
\begin{equation}
N=6\frac{M_{t}^{2}}{M_{H}^{2}}-6\frac{M_{t}^{2}}{M_{H}^{2}}\left(
4\frac{M_{t}^{2}}{M_{H}^{2}}-1\right)  \left(  \arcsin\frac{M_{H}}{2M_{t}%
}\right)  ^{2}. \label{Higgs_N}%
\end{equation}

It turns out that the structure of the loop integral over $Q_{\intercal}$ has
exactly the same form like that for gluon jets case (\ref{constR}). So to
describe the exclusive Higgs production we take the result (\ref{H-pp}) and
replace $R^{2}=1$ by $\tilde{R}^{2}\left(  \mu\right)  $ with $\tilde
{R}\left(  \mu\right)  $ given by the formula (\ref{R-tylda}), where
$\mu=M_{gg}/2\rightarrow M_{H}/2$.

Performing the appropriate calculations we find the differential cross section
$d\sigma/dy$ (not presented in \cite{Bial-Land}) for DPE exclusive Higgs boson
production to be in the form:
\begin{equation}
\frac{d\sigma}{dy}=\frac{BN^{2}}{4^{5}\pi^{3}\alpha^{\prime}{}^{2}}\left(
\frac{s}{M_{H}^{2}}\right)  ^{2\epsilon}\frac{\tilde{R}^{2}\left(
M_{H}/2\right)  }{\left(  (\lambda+\beta)/\alpha^{\prime}-\ln\left[
M_{H}/\sqrt{s}\right]  \right)  ^{2}-y^{2}}. \label{Higgs_y}%
\end{equation}
Here $y$, $\ln(M_{H}/\left(  \delta_{2}^{\max}\sqrt{s}\right)  )\leqslant
y\leqslant\ln(\delta_{1}^{\max}\sqrt{s}/M_{H})$, is a rapidity of the produced
Higgs. In the following we use $\delta_{1}^{\max}=\delta_{2}^{\max}%
=\delta=0.1$. Since $y^{\max}=\ln(\delta\sqrt{s}/M_{H})$ and $(\lambda
+\beta)/\alpha^{\prime}=12$ the differential cross section (\ref{Higgs_y})
very weekly depends on the rapidity of the Higgs. So to get the total cross
section for DPE exclusive Higgs production it is enough to multiply the above
result (\ref{Higgs_y}) (at $y=0$) by a factor $y^{\max}-y^{\min}=$ $\ln\left(
s\delta^{2}/M_{H}^{2}\right)  $ what leads to the final result:
\begin{equation}
\sigma=\frac{BN^{2}}{4^{5}\pi^{3}\alpha^{\prime}{}^{2}}\left(  \frac{s}%
{M_{H}^{2}}\right)  ^{2\epsilon}\frac{\tilde{R}^{2}\left(  M_{H}/2\right)
}{\left(  (\lambda+\beta)/\alpha^{\prime}-\ln\left[  M_{H}/\sqrt{s}\right]
\right)  ^{2}}\ln\left(  \frac{s\delta^{2}}{M_{H}^{2}}\right)  .
\label{Higgs_t}%
\end{equation}

Now we are ready to give our predictions for DPE exclusive Higgs production at
the Tevatron and the LHC energies. We also compare our results with those
obtained in a model developed by Khoze, Martin and Ryskin (KMR model).

In Table \ref{Higgs_TeV} the prediction for the Tevatron energy, $\sqrt{s}=2$
TeV, is shown. The mass of the Higgs is taken to be $120$ GeV and $\alpha
_{s}(M_{H})$ is about $0.1$. We also include the $\alpha_{s}$ virtual
correction \cite{Khoze-Higgs-Jets,K-factor} to the $gg\rightarrow H$ vertex
factor, so-called $K$-factor to be about $1.5$. As was discussed earlier we
take $\delta=0.1$ and assume $S_{\text{gap}}^{2}/\left(  G^{2}/4\pi\right)
^{2}$ to be $0.6$.\begin{table}[h]
\begin{center}%
\begin{tabular}
[c]{|c|c|c|}\hline\hline
$\sqrt{s}=2$ TeV & $%
\begin{array}
[c]{c}%
\text{KMR model}\\
S_{\text{gap}}^{2}=0.05
\end{array}
$ & $%
\begin{array}
[c]{c}%
\text{Our model}\\
S_{\text{gap}}^{2}/\left(  G^{2}/4\pi\right)  ^{2}=0.6
\end{array}
$\\\hline
$\sigma$ [fb] & $0.06$ & $0.005$\\\hline
\end{tabular}
\end{center}
\caption{Our result for DPE exclusive Higgs production cross section for the
Tevatron energy. Our prediction is about $10$ times smaller than the
prediction based on the KMR model. }%
\label{Higgs_TeV}%
\end{table}

Before we present the prediction for the LHC energy, $\sqrt{s}=14$ TeV, we
have to take into account the $s$ dependence of the gap survival factor
$S_{\text{gap}}^{2}$. Following \cite{S2-theory} we expect that $S_{\text{gap}%
}^{2}\left(  \sqrt{s}=2\text{ TeV}\right)  /S_{\text{gap}}^{2}\left(  \sqrt
{s}=14\text{ TeV}\right)  \approx0.4$ what allows us to assume $S_{\text{gap}%
}^{2}/\left(  G^{2}/4\pi\right)  ^{2}=0.25$. The obtained result for DPE
exclusive Higgs ($M_{H}=120$ GeV) production at the LHC energy is presented in
Table \ref{Higgs_LHC}.\begin{table}[h]
\begin{center}%
\begin{tabular}
[c]{|c|c|c|}\hline\hline
$\sqrt{s}=14$ TeV & $%
\begin{array}
[c]{c}%
\text{KMR model}\\
S_{\text{gap}}^{2}=0.02
\end{array}
$ & $%
\begin{array}
[c]{c}%
\text{Our model}\\
S_{\text{gap}}^{2}/\left(  G^{2}/4\pi\right)  ^{2}=0.25
\end{array}
$\\\hline
$\sigma$ [fb] & $2$ & $0.015$\\\hline
\end{tabular}
\end{center}
\caption{Our result for DPE exclusive Higgs production cross section for the
LHC energy. A distinct difference, $\sim10^{2}$, with the KMR model prediction
is observed. }%
\label{Higgs_LHC}%
\end{table}

As can be seen from Table \ref{Higgs_TeV} and Table \ref{Higgs_LHC} the
results for DPE exclusive Higgs production are about one order of magnitude
smaller than those obtained in the KMR model
\cite{Khoze-Higgs-Jets,Khoze-Higgs} for the Tevatron energy and about two
orders of magnitude smaller for the LHC energy. It reflects the possible large
uncertainties of the presented approach and the general fact that the
perturbative QCD predictions, on the contrary to the non-perturbative
two-gluon-exchange-type models, show a strong increase of the cross sections
with increasing energy. Hopefully a study of the dijets production as a
function of the energy will clearly be able to discriminate between the
perturbative QCD determinations and the non-perturbative model approaches.

\bigskip

\bigskip

\bigskip

\textbf{Acknowledgements }It is a pleasure to thank Dr. Leszek Motyka, Prof.
Andrzej Bialas and Prof. Robert Peschanski for useful discussions. This
investigation was supported by the Polish State Committee for Scientific
Research (KBN) under grant 2 P03B 043 24.

\end{document}